\documentclass{article}
\usepackage{PRIMEarxiv}

\usepackage[utf8]{inputenc} 
\usepackage[T1]{fontenc}    
\usepackage{hyperref}       
\usepackage{url}            
\usepackage{booktabs}       
\usepackage{amsfonts}       
\usepackage{nicefrac}       
\usepackage{microtype}      
\usepackage{lipsum}
\usepackage{fancyhdr}       
\usepackage{graphicx}       
\graphicspath{{media/}}     
\usepackage{amsmath,bm}
\usepackage{hyperref}       
\usepackage{cleveref}
\usepackage{amsfonts}       
\usepackage{graphicx}
\usepackage{multicol}
\usepackage{multirow}
\usepackage{float}
\usepackage{listings}
\usepackage[english]{babel}
\usepackage{amsthm}
\usepackage{caption}
\usepackage{subcaption}
\usepackage{xcolor}
\usepackage{mathtools}
\usepackage{authblk}

\title{Large Transformers are Better EEG Learners}

\author[1]{\small Bingxin WANG}
\author[1]{\small Xiaowen Fu}
\author[1]{\small Yuan LAN}
\author[3*]{\small Luchan ZHANG}
\author[4*]{\small Wei ZHENG}
\author[1,2*]{\small Yang Xiang}

\affil[1]{\footnotesize Department of Mathematics, The Hong Kong University of Science and Technology, Clear Water Bay, Hong Kong SAR, China}
\affil[2]{\footnotesize Algorithms of Machine Learning and Autonomous Driving Research Lab, HKUST Shenzhen-Hong Kong Collaborative Innovation Research Institute, Futian, Shenzhen, China}
\affil[3]{\footnotesize College of Mathematics and Statistics, Shenzhen University, Shenzhen 518060, China}
\affil[4]{\footnotesize Shenzhen Youjia Innov Tech Co., Ltd, Shenzhen, China}
\affil[*]{Corresponding authors:  zhengwei@minieye.cc, zhanglc@szu.edu.cn, maxiang@ust.hk}

\begin{document}
\maketitle
\begin{abstract}
Pre-trained large transformer models have achieved remarkable performance in the fields of natural language processing and computer vision. 
However, the limited availability of public electroencephalogram (EEG) data presents a unique challenge for extending the success of these models to EEG-based tasks. 
To address this gap, we propose AdaCT, plug-and-play \textbf{Ada}pters designed for \textbf{C}onverting \textbf{T}ime series data into spatio-temporal 2D pseudo-images or text forms. Essentially, AdaCT-I transforms multi-channel or lengthy single-channel time series data into spatio-temporal 2D pseudo-images for fine-tuning pre-trained vision transformers, while AdaCT-T converts short single-channel data into text for fine-tuning pre-trained language transformers. The proposed approach allows for seamless integration of pre-trained vision models and language models in time series decoding tasks, particularly in EEG data analysis. Experimental results on diverse benchmark datasets, including Epileptic Seizure Recognition, Sleep-EDF, and UCI HAR, demonstrate the superiority of AdaCT over baseline methods.  
Overall, we provide a promising transfer learning framework for leveraging the capabilities of pre-trained vision and language models in EEG-based tasks, thereby advancing the field of time series decoding and enhancing interpretability in EEG data analysis.
Our code will be available at https://github.com/wangbxj1234/AdaCE.
\end{abstract}

\section{Introduction}\label{sec:introduction}
Electroencephalography (EEG) has long been instrumental in unraveling the intricacies of the human brain.
In contrast to text, audio, or video, modeling EEG data presents three main challenging characteristics.
Firstly, the magnitude of available public electroencephalogram (EEG) data is significantly lower than that of text and image data. This poses a unique obstacle for transformer models pre-trained from EEG data, making it nearly impossible to reach the scale of other large transformers.
Secondly, multi-channel EEG data analysis for predictive and classificatory purposes requires a comprehensive understanding of both its temporal and spatial dimensions. The temporal aspect refers to the evolution of neuroelectric signals over time, while the spatial dimension involves the distribution of these signals across multiple channels. Extracting meaningful information from EEG data relies on interpreting its inherent texture, which is manifested in the time-spatial dependency derived from multi-channel recordings. 
Thirdly, in most existing EEG prediction or classification tasks, the primary objective is to capture immediate stimuli or distinctive pre-ailment patterns from lengthy time series data. However, a significant challenge arises due to the inherent quadratic complexity of attention computation, imposing limitations on the input length of transformer-based models. This presents a hurdle in transforming the raw EEG data, with its extended sequences, into a format suitable for the model without compromising information integrity. 

\begin{figure*}
\centering
\includegraphics[width=0.8\textwidth]{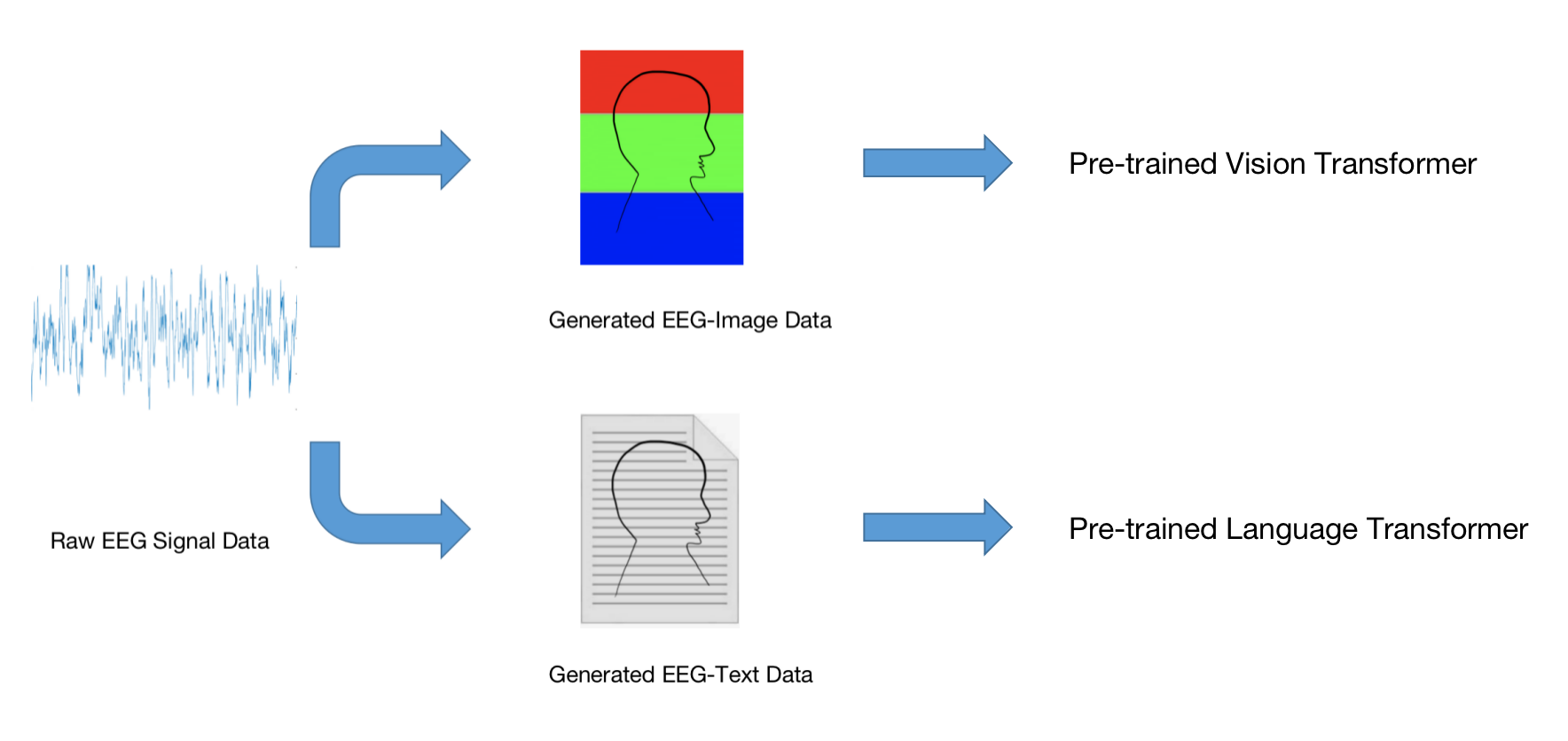} 
\caption{Framework: Adapters for converting time series EEG data into images or text for fine-tuning pre-trained large transformers.}
\label{fig1}
\end{figure*}

To bridge the gap between the scarcity of available EEG data and the potential of large transformer models pretrained on other modalities, in this paper, we demonstrate that large models (LMs) pre-trained from images as well as text can be fine-tuned for EEG-based prediction tasks without introducing extra parameters to be trained.
In particular, we introduce AdaCT, plug-and-play Adapters designed for
converting time series data into spatio-temporal 2D pseudo-images or text, to apply pretrained LMs for EEG prediction problems, with the framework illustrated in Fig. \ref{fig1}. 
Fundamentally, this method converts multi-channel or lengthy single-channel time series data into spatio-temporal 2D pseudo-images, while converting short single-channel data into text.

To achieve robust performance, we propose a three-stage method for converting EEG data into pseudo-images compatible with pre-trained visual transformers while maximizing the retention of pertinent information.
Leveraging the enhanced generalization capabilities acquired through pre-training, we show how pre-trained vision transformers can capture the complete texture features embedded in the spatio-temporal EEG pseudo-images. 
For adapting short single-channel EEG to pre-trained language transformer models, we convert the temporal dynamics of the EEG signal into a text-based representation, leveraging the inherent sequential nature of both the EEG data and text. 

AdaCT seamlessly integrates with pre-trained transformers, facilitating the application of cutting-edge models to EEG prediction without introducing extra parameters. Our evaluations reveal AdaCT's ability to surpass or match state-of-the-art EEG prediction methods. Furthermore, it achieves top-tier results on other multi-channel human activity recording time series datasets, including the UCI HAR benchmark\cite{anguita2013public}.

\section{Related Works}\label{sec:Related Works}
\subsection{Transformer-Based EEG Decoding Methods}

In the area of EEG decoding, transformer-based models have been leveraged for enhancing the capture of long-term dependencies in studies such as those conducted by ~\cite{sun2021eeg}, ~\cite{9845479}, and ~\cite{song2022eeg}. Each of these investigations significantly contributed to advancing this research paradigm. A common methodology across these studies involved the incorporation of a combined Convolutional Neural Network (CNN) module. This module, executing convolutions separately across the dimensions of time and space, facilitated the pre-extraction of both temporal and spatial features.
These approaches collectively represent a nuanced strategy for empowering models to discern intricate patterns in EEG data, thereby paving the way for advancements in the understanding and decoding of complex neural signals.

However, introducing a CNN pre-extraction module to pre-trained large transformers can pose challenges. This addition introduces additional parameters that require training. During the fine-tuning process, backpropagating the gradients of these parameters throughout the entire model significantly increases computational costs.

\subsection{BERT-like Transformers Pre-trained on EEG Datasets}

The pioneering studies conducted by ~\cite{kostas2021bendr} and ~\cite{nogales2022bert} have set the stage for a transformative approach in EEG decoding. These investigations stand out for their ingenious utilization of pre-trained BERT-like transformers, marking a commendable leap forward in the convergence of natural language processing and neuroscientific research. Leveraging pre-trained language models to decode EEG data represents a novel and promising avenue, showcasing the adaptability and potential of transformer technology in a field traditionally dominated by more conventional methods.

However, this innovative approach encounters inherent challenges. The scalability of pre-trained transformers on EEG datasets is constrained by the limited availability of public EEG data compared to more abundant text and image datasets. Furthermore, the application of large language transformers faces hurdles related to managing input sequence length. The quadratic complexity of attention calculation imposes limitations on the allowable length of input sequences, with challenges arising particularly when attempting to consolidate multiple channels into a unified text input.

\subsection{Generative Pre-trained Transformer (GPT)}

Generative Pre-trained Transformer (GPT) ~\cite{radford2018improving} models, pre-trained on vast textual corpora, have garnered acclaim in natural language processing tasks. Their effectiveness lies in their ability to grasp contextual dependencies and generate coherent text. This pre-training enables GPT to capture intricate linguistic patterns and nuances, making it a powerful tool for language-related applications. The adaptation of pre-trained GPT models in fields beyond language processing is an ongoing exploration. Our proposed transfer learning framework contributes to this exploration by facilitating their application in EEG decoding tasks, where the pre-trained models demonstrate promising generalization capabilities.

\subsection{Pre-trained Vision Transformer (ViT)}

Vision Transformer, introduced by ~\cite{dosovitskiy2020image}, revolutionized computer vision by applying transformer architectures directly to image data. Unlike traditional convolutional neural networks (CNNs), ViT abandons convolutions in favor of self-attention mechanisms. The image is divided into fixed-size patches, linearly embedded, and processed by transformer layers, enabling the model to capture both local and global features. By pre-training on extensive image datasets, ViT learns rich visual representations and global dependencies. This pre-trained knowledge, when fine-tuned for specific visual tasks, accelerates convergence and demonstrates superior performance in discerning local and global patterns. 
Based on the insights discussed above, we believe that once a reasonable method is found to convert the time series data into pseudo-images, the pre-trained ViT model can serve as an effective feature extractor for EEG classification tasks.

\begin{figure*}
\centering
\includegraphics[width=1\textwidth]{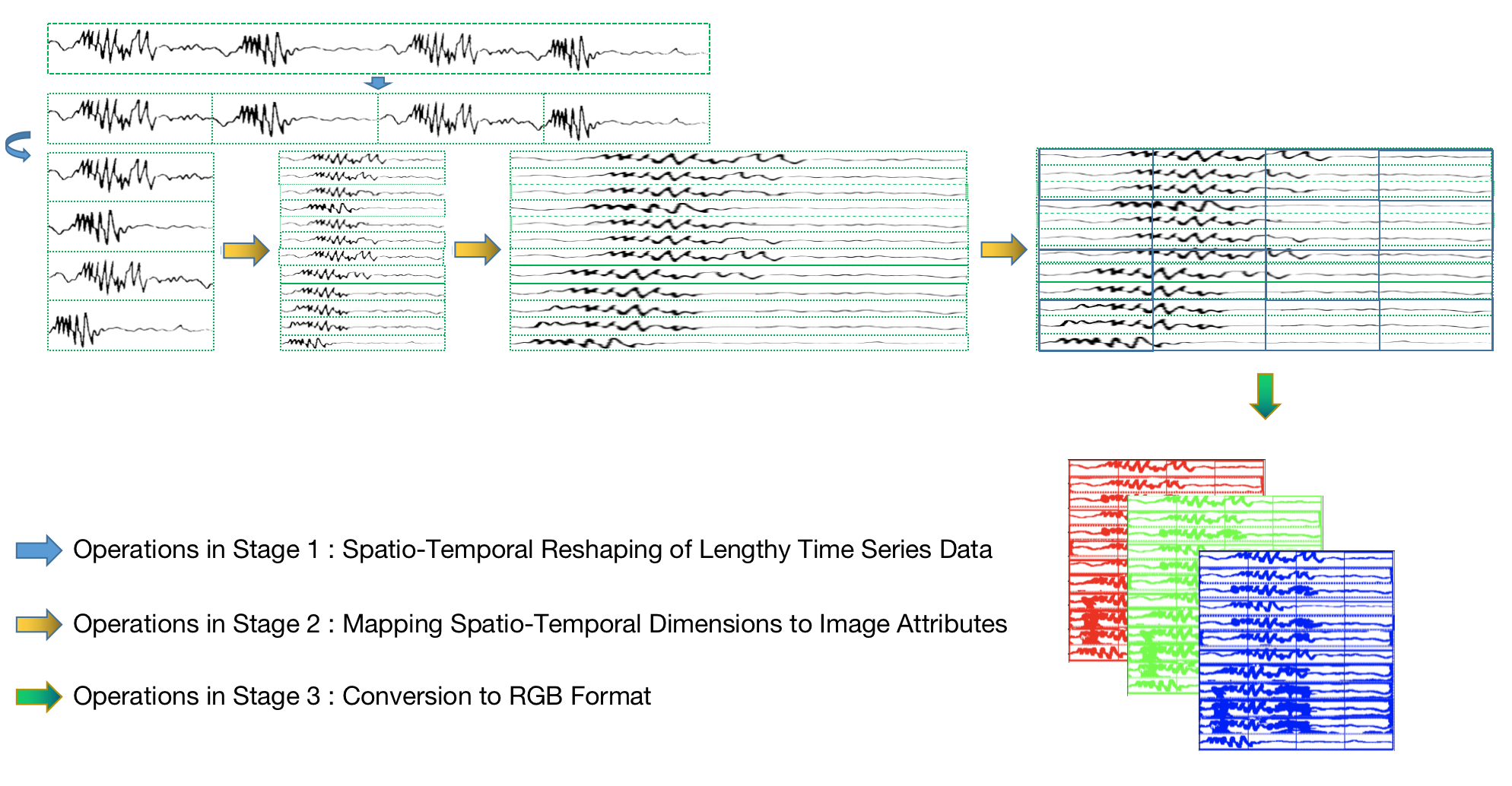} 
\caption{Illustration of the AdaCT-I method, showcasing the spatio-temporal reshaping, mapping to image attributes, and conversion to RGB format steps for converting time series data into two-dimensional RGB images.}
\label{fig2}
\end{figure*}

\begin{figure}[t]
\centering
\includegraphics[width=0.6\columnwidth]{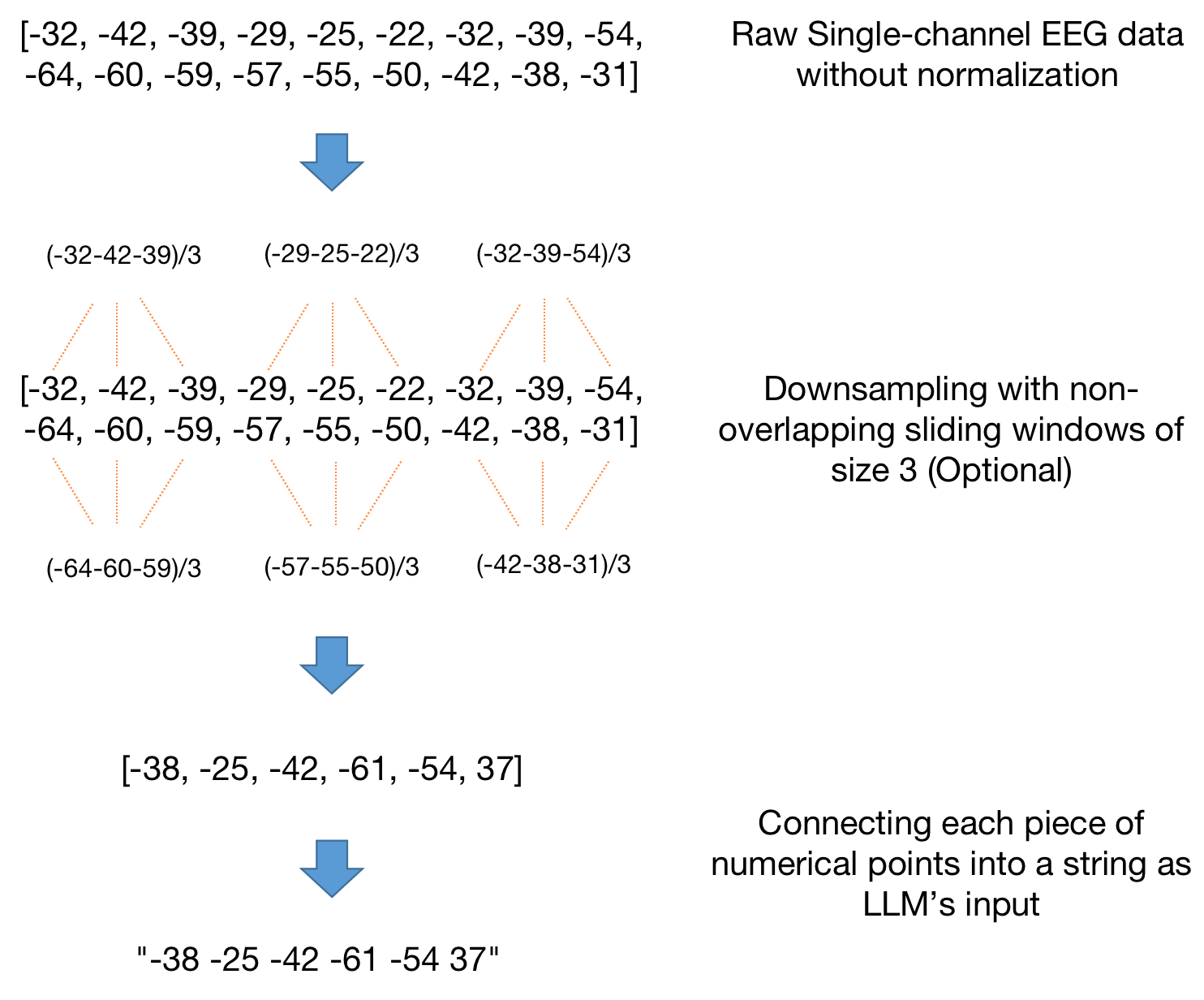} 
\caption{Illustration of the AdaCT-T method, highlighting the non-overlapping sliding window downsampling step for converting time series data into text representation.}
\label{fig3}
\end{figure} 

\begin{figure*}
\centering
\includegraphics[width=1\textwidth]{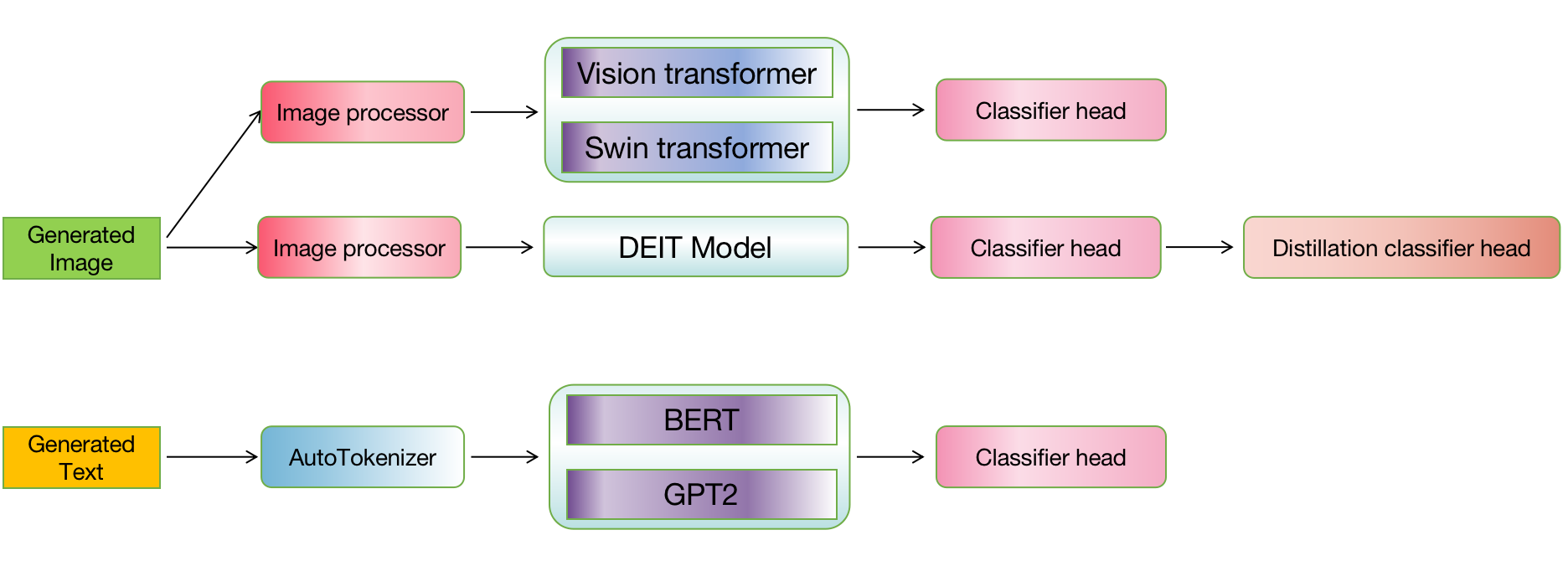} 
\caption{Overview of the fine-tuning process for pre-trained vision transformers and language transformers on converted EEG datasets. The process involves image processing for vision transformers and tokenization for language transformers, followed by integration with pre-trained models and classification head modules.}
\label{fig4}
\end{figure*}

\section{AdaCT: Adapters for Converting Time Series Data into Images or Text}\label{sec:method}
In this section, we delineate our solution crafted for fine-tuning pre-trained transformers on EEG datasets. To enhance its applicability, we have developed plug-and-play adapters designed for transforming raw EEG data into a format directly compatible with general pre-trained transformer models. 
The purpose of this approach is to leverage the generalization capabilities inherent in transformer models pre-trained on diverse datasets to improve the performance of EEG prediction and classification tasks.

Generally, with the aim of more effectively harnessing the diverse characteristics of pre-trained visual and language transformers, we seek to partition distinct transformation approaches based on the length and channel count of the data. Subsequently, the following sections intricately expound upon the specific implementations of the proposed adapters, namely AdaCT-I and AdaCT-T. AdaCT-I concentrates on the conversion of multi-channel or lengthy EEG data into spatio-temporal 2D pseudo-images. Simultaneously, AdaCT-T investigates the adjustment of pre-trained language transformers for short single-channel EEG, translating temporal dynamics into a text-based representation.

\subsection{AdaCT-I: Adapt Time Series Data into Images}

Our goal is to convert raw EEG data into two-dimensional RGB images, emphasizing the preservation of complete texture characteristics. The adaptation process includes three stages, as shown in Fig. \ref{fig2}.

\subsubsection{Spatio-Temporal Reshaping of Lengthy Time Series Data}

Handling lengthy time series introduces challenges, such as potential information loss and model input constraints. In addressing these challenges, we found that the core of most EEG prediction or classification tasks is to capture immediate stimuli or distinctive pre-ailment patterns, in which the contribution of long temporal dependencies is insignificant.  Based on this insight, we propose partitioning the signal into evenly spaced patches along the temporal axis and stacking them spatially, augmenting the spatial dimensionality. Instead of focusing on preserving extended temporal correlations, our approach prioritizes essential features for effective modeling. For instance, in the case of single-channel EEG data with a time series length of 65536, we treat every 256 points as a patch, reshaping it into 256x256 spatio-temporal data. This pre-adjustment ensures the model effectively captures critical information without being burdened by unnecessary temporal intricacies, especially in scenarios with limited channels and extended time durations.

\subsubsection{Mapping Spatio-Temporal Dimensions to Image Attributes}

Transforming multi-channel time series EEG data into a 2D pseudo-image centers on establishing a direct mapping from the sampling points to image pixels. In our approach, each sampling point in the time series corresponds to a pixel, and linear interpolation is applied to emphasize channel data importance and enhance inter-channel correlation. This ensures the preservation of essential spatio-temporal features critical for subsequent tasks.

\subsubsection{Conversion to RGB Format}

In the final stage, we convert the 2D pseudo-images into RGB format, ensuring seamless compatibility with the input interfaces of most pre-trained visual transformers. Notably, vision transformers will flatten the RGB channels of input images before the attention calculation, for example, flattening (16, 16, 3) into a vector of size (16*16*3). 
To preserve the original structure and content, we apply the inverse operation of flattening, namely folding, to convert the 2D pseudo-images into RGB format. 
This folding operation allows us to reconstruct the initial 2D pseudo-images from the flattened representation without any kind of information loss.

\subsection{AdaCT-T: Adapt Time Series Data into Text}
\label{3.2}

To adapt pre-trained transformer models to single-channel EEG data, we propose a straightforward yet effective approach. We convert the temporal dynamics of the EEG signal into a text-based representation, leveraging the inherent sequential nature of both the EEG data and text. This involves intuitively connecting sequential sampling points with spaces, creating a coherent text string. Prior to implementing this mapping, we conduct a two-stage process to optimize the input format for subsequent model utilization.

\subsubsection{Temporal Data Scaling for Signal Representation}
For a given dataset, following an examination of the overall data distribution, we employ a consistent scaling method to convert each temporal data point of the EEG signal into a three-digit integer. It is important to note that, although datasets may showcase diverse distributions, data within the same dataset typically adheres to consistent upper and lower limits. Thus, to guarantee a succinct numeric representation for each temporal data point, we adapt scaling methods based on the unique characteristics of each dataset. As an illustration, for a normalized dataset, we utilize a scaling approach that involves multiplying by a thousand and subsequently rounding the result.

\subsubsection{Non-overlapping Sliding Window Downsampling}
When addressing EEG data, the densely sampled nature introduces pronounced continuity characteristics among neighboring sampling points on the timestamp. Exploiting this feature, we apply a non-overlapping sliding window downsampling technique, adjusting the data to meet the maximum input length of the target language transformer, as illustrated in Fig. \ref{fig3}.

The crux of this approach lies in employing non-overlapping sliding windows: within each window, the average of sampling points is computed for downsampling. It is crucial to emphasize that we deliberately choose a relatively small window size (typically set to 3 in most experiments). This strategic choice proves beneficial as it condenses extensive time series data while preserving fundamental temporal texture information.

\subsection{Fine-tune Pre-trained Vision Transformers on Converted Datasets}
Following the transformation of EEG, we align the generated image and text datasets with the transformer model's input requirements for their utilization in fine-tuning pre-trained transformers. For AdaCT-I, the pre-folded spatio-temporal dimensions of the 2D pseudo-images undergo standard patch cutting processes, dividing them into smaller patches for the transformer to efficiently process localized information. Concurrently, in AdaCT-T, tokenization is applied to the text-based representation of EEG data. The sequential nature of the text format aligns seamlessly with the transformer's capacity to capture dependencies among tokens, facilitating effective training. The overall framework is depicted in Fig. \ref{fig4}.

Fine-tuning pre-trained vision transformer models for EEG data processing involves a sequence of steps. We provide a detailed explanation of the architecture in three modules:

\subsubsection{Image Processing Module}

The objective of the image processing module is to prepare the pseudo-images of EEG data for compatibility with pre-trained vision transformer models. This transformation is accomplished through the following operations:

\begin{itemize}
    \item \textbf{Resizing and Standardization}: The image processor firstly resizes the input images to a fixed size, ensuring uniformity and compatibility with the requirements of the vision transformer models. Additionally, color channels are standardized, and pixel values are normalized to promote uniformity across different images.

    \item \textbf{Normalization}: Following resizing, normalization techniques are applied to standardize the pixel values of the input images to a predefined scale, typically ranging from 0 to 1 or -1 to 1. This normalization step stabilizes the training process and enhances the convergence of vision transformer models during subsequent fine-tuning.

    \item \textbf{Augmentation and Encoding}: Subsequently, augmentation techniques are employed to enhance the diversity of the converted training data artificially. Techniques such as rotation, flipping, and cropping contribute to improving the robustness and generalization capabilities of the vision transformer models. Finally, the preprocessed images are encoded into numerical tensors or vectors, preserving both spatial and semantic information, thus rendering them suitable for input into the vision transformer models.

\end{itemize}

The image processing module ensures the effective transformation of input images into a standardized format encapsulating the essential visual features and semantics of EEG data. This standardized representation facilitates subsequent processing by transformer models, enabling efficient extraction of meaningful insights from the input data. Ultimately, these enhancements contribute to the improved performance and accuracy of EEG-related tasks such as prediction and classification.

\subsubsection{Integration with Pre-trained Vision Transformer Models}

After being processed by the image processor module, the preprocessed images are primed for direct integration into cutting-edge vision transformer models to extract features. In this paper, we select three popular vision transformer architectures to serve as feature extractors for the pseudo-images generated by AdaCT-I: ViT~\cite{dosovitskiy2020image}, Swin Transformer~\cite{DBLP:journals/corr/abs-2103-14030}, and DeiT ~\cite{touvron2021training}.
The integration of preprocessed EEG pseudo-images with vision transformer models is crucial to ensuring high-precision prediction. In the following discussion, we analyze the characteristics of three vision models and their suitability for processing EEG-derived data.

\begin{itemize}
\item \textbf{Vision Transformer (ViT)}: In the architecture of ViT~\cite{dosovitskiy2020image}, the input image is divided into fixed-size patches, linearly embedded, and processed by transformer layers.  Moreover, it is often observed that immediate stimuli or distinctive pre-ailment patterns are represented by only a small subset of adjacent original sampling points. Therefore, through the spatio-temporal reshaping of EEG data by AdaCT-I and the patch-wise partitioning approach of ViT, most of the critical texture information remains intact within each patch. By capitalizing on its inherent ability to capture spatial and temporal patterns effectively, ViT emerges as a robust choice for our task. Its architecture is adept at handling the nuanced complexities inherent in the transformed EEG data, thus facilitating the extraction of meaningful features for subsequent analysis.

\item \textbf{Swin Transformer}: The Swin Transformer~\cite{DBLP:journals/corr/abs-2103-14030} incorporates shifted windows, enabling efficient capture of correlated features across different windows compared to ViT. While ViT generally retains most waveforms containing key EEG texture information within the same patch, in some instances, the required texture is dispersed across multiple patches after partitioning.  
The window shifting technique effectively addresses this issue, mitigating information loss and facilitating comprehensive feature extraction across various spatial contexts. Moreover, the Swin Transformer's hierarchical architecture facilitates feature aggregation across multiple scales, allowing it to extract both local and global information from the input data. These characteristics enable the Swin Transformer model to extract features effectively from the EEG-derived data.

\item \textbf{Data-efficient Image Transformer (DeiT)}: The DeiT model~\cite{touvron2021training} is a distilled Vision Transformer that utilizes a distillation token to learn from a CNN teacher during pre-training on conventional image datasets. This approach contributes to enhanced generalization of the model, enabling it to transfer its image feature extraction capabilities to datasets generated by AdaCT-I. Additionally, DeiT's architecture is specifically designed to efficiently process image data and capture both local and global features, making it  well-suited for learning the inherent texture with the transformed EEG images.  

\end{itemize}

The vision models we used were pre-trained on ImageNet-1k or ImageNet-21k~\cite{deng2009imagenet}, leveraging the benefits of pre-training to learn rich visual representations and global dependencies. Pre-training on extensive image datasets enables the models to acquire a broad understanding of visual patterns and structures, enhancing their capability to extract meaningful features from input data. By fine-tuning these pre-trained models on EEG data processing, we capitalize on the pre-learned knowledge to accelerate convergence and improve performance in discerning both local and global patterns in the transformed EEG images. 

\subsubsection{Classifier Head Module}

In the final module, the extracted features from the transformer model are passed into a classifier head. This linear classification head transforms the encoded representations produced by the transformer's self-attention mechanism into predictions for specific tasks, mapping them to the output space required for EEG-related tasks such as prediction and classification.

Particularly, the DeiT model requires an additional distillation classifier head after the primary classifier head. This supplementary head is integral during the pre-training process, as it distills knowledge from the CNN teacher, thereby enhancing the model's performance by leveraging the pre-trained knowledge.
We keep the additional distillation classifier structure during fine-tuning of the DeiT model for leveraging its pre-learned features.

\subsection{Fine-tune Self-Supervised Pre-trained Language Transformers on Converted Datasets}
In this section, we explain the fine-tuning process of pre-trained language transformers on the text-based representation of EEG data. The text-based format of EEG data, generated by AdaCT-T, serves as the input for language transformers. We introduce three key modules utilized in this fine-tuning process: the AutoTokenizer module, GPT-2/BERT model integration module and the classification head module.

\subsubsection{AutoTokenizer Module}

The AutoTokenizer module is designed to prepare the text generated from EEG data for compatibility with pre-trained language transformer models. This transformation is accomplished through the following operations:

\begin{itemize}
\item \textbf{Tokenization and Special Token Handling}: This initial step involves breaking down input text into tokens while incorporating special tokens like [CLS], [SEP], and [MASK] for sequence delimiting and masking purposes.

\item \textbf{Subword Tokenization and Vocabulary Mapping}: Subsequently, words are decomposed into subword units, enhancing vocabulary coverage. Tokens are then mapped to numerical representations based on the model's vocabulary.

\item \textbf{Padding and Attention Masking}: The module ensures uniform input sequence length by padding shorter sequences. Attention masking mechanisms are applied to focus model attention on relevant segments, disregarding padding tokens.

\item \textbf{Normalization and Encoding}: Following tokenization, normalization techniques standardize pixel values to predefined scales. Augmentation methods diversify training data, and images are encoded into numerical tensors, preserving spatial and semantic information.

\end{itemize}

The AutoTokenizer module serves as a vital bridge between the converted textual data and transformer architectures, optimizing text preprocessing for efficient model utilization in the EEG-related tasks.

\subsubsection{Integration with Self-Supervised Pre-trained Language Transformer Models}

The integration of preprocessed textual data with language transformer models is imperative for ensuring accurate predictions. In this study, we use two language transformer architectures, BERT~\cite{DBLP:journals/corr/abs-1810-04805} and GPT-2~\cite{radford2019language}, to serve as feature extractors for the text generated by AdaCT-T:

\begin{itemize} 
\item \textbf{BERT (Bidirectional Encoder Representations from Transformers)}: The bidirectional attention mechanism in BERT~\cite{DBLP:journals/corr/abs-1810-04805} enhances the model's ability to comprehend intricate linguistic nuances and dependencies within the input text. When processing textual data generated from EEG signals, we suppose this capability will be instrumental in capturing the subtle variations in the inherent texture of EEG signals. Consequently, it enables the model to effectively capture immediate stimuli or distinctive pre-ailment patterns lurking within the lengthy strings of time series data.

\item \textbf{GPT-2 (Generative Pre-trained Transformer 2)}: Different from BERT's bidirectional approach, GPT-2~\cite{radford2019language} employs a unidirectional transformer architecture, processing text sequentially from left to right. Apart from its unidirectional processing, GPT-2 leverages an autoregressive decoding mechanism, enabling it to generate coherent and contextually relevant text. Despite its generative nature, GPT-2 effectively captures nuanced textual features essential for classification tasks through its autoregressive decoding and contextual embeddings. This capability allows GPT-2 to analyze and extract relevant features from the processed EEG text, contributing to its adaptability for classification purposes.

\end{itemize}
 
Both BERT and GPT-2 models are pre-trained on large corpora of text data using self-supervised learning techniques, which enable them to learn rich linguistic representations and capture intricate semantic relationships. Self-supervised learning involves training models to predict missing or masked portions of input data, without relying on explicit supervision or labeled examples. This process encourages the models to learn meaningful features and structures from the raw text data, leading to highly effective representations. By fine-tuning these pre-trained models on EEG-derived textual data, we leverage their pre-learned knowledge obtained through self-supervised learning to expedite convergence and enhance performance in EEG-related tasks such as prediction and classification.

\subsubsection{Classifier Head Module}

Following the paradigm established in the discussion of vision models, the classifier head for the language transformer operates in a similar way. It receives the extracted features from the transformer model and transforms them into predictions by maps them to the output space required for EEG-related tasks.

\begin{figure*}
\centering
\includegraphics[width=1\columnwidth]{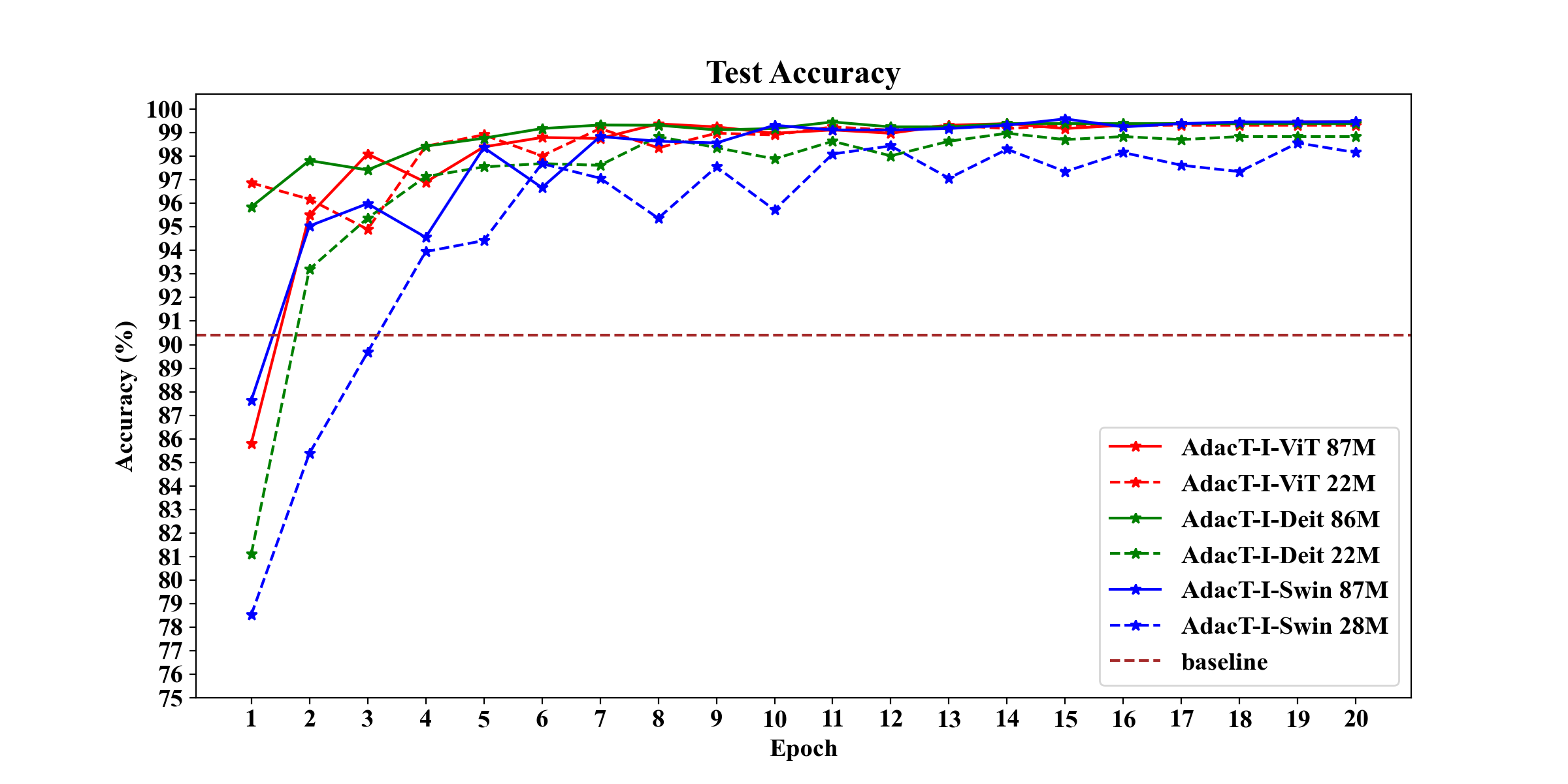} 
\caption{Fine-tuning Process: Epoch-wise Comparative Analysis of AdaCT-I on UCI HAR Dataset Using Various Pre-trained Vision Models with Baseline (TS-TCC~\cite{eldele2021time}) Accuracy.}
\label{fig5}
\end{figure*}

\section{Experiments}\label{sec:exp}
\subsection{Experiment Details}
We apply AdaCT to diverse EEG datasets and the UCI HAR dataset~\cite{anguita2013public}, guiding the fine-tuning process from the smallest to the larger pre-trained transformers. Our quantitative evaluation, involving calculations for average accuracy and macro-F1 score, consistently reveals compelling results. Notably, fine-tuning transformers with only 100 million to 300 million parameters on AdaCT-converted datasets could match or surpass benchmarks established by previous state-of-the-art methods. These findings underscore AdaCT's effectiveness in enhancing the adaptability of pre-trained transformers, particularly for EEG and some other multi-channel human activity recording time series data.

Specifically, we evaluate our method on three datasets: Epileptic Seizure Recognition~\cite{misc_epileptic_seizure_recognition_388}, Sleep-EDF~\cite{goldberger2000physiobank}, and UCI HAR~\cite{anguita2013public}. 
Following the setting in ~\cite{eldele2021time}, we split the datasets into 60\%, 20\%, and 20\% for training, validating, and testing. 
We fine-tune the pre-trained transformer models using Huggingface Trainer~\cite{wolf2019huggingface} with AdamW optimizer, initial learning rate of 5e-5, training batch size per device of 16, and gradient accumulation steps of 4.
We report the prediction accuracy and macro-averaged f1 scores.
Each experiment we conduct is trained within 20 epochs, and further training did not significantly improve the performance.
Our training environment is PyTorch 1.13.0+cu11 on two NVIDIA Quadro RTX 8000 GPUs.

\subsection{Datasets}

\subsubsection{Epileptic Seizure Prediction}
The Epileptic Seizure Dataset~\cite{misc_epileptic_seizure_recognition_388} consists of 500 files, with each file representing a single subject. Each file is a recording of brain activity for 23.6 seconds. The corresponding time series was sampled into 4097 data points. Every 4097 data points were divided and shuffled into 23 chunks, and each chunk contains 178 data points for 1 second. There are 23 x 500 = 11500 pieces of information.

\subsubsection{Sleep-EDF}

The Sleep-EDF Dataset~\cite{goldberger2000physiobank} contains 197 whole-night PolySomnoGraphic sleep recordings, containing EEG, EOG, chin EMG, and event markers.
We follow ~\cite{eldele2021attention}, using the first 20 subjects' records out of 78 to construct the single-channel EEG train dataset of 42308 pieces of information, with each piece of information containing 1*3000 data points.

\subsubsection{UCI HAR}
The Human Activity Recognition Dataset~\cite{anguita2013public} was collected from 30 subjects performing six different activities (Walking, Walking Upstairs, Walking Downstairs, Sitting, Standing, Laying), consisting of inertial sensor data that was collected using a smartphone carried by the subjects.
The inside EEG data has nine channels: three for acceleration signals of X, Y, and Z; three for body acceleration obtained by subtracting gravity from the total acceleration of X, Y, and Z; three for the angular velocity vector of X, Y, and Z.
There are 7352 pieces of EEG data, each piece containing 9*128 data points.


\begin{table}
\centering
\begin{tabular}{|l|l|l|}
    \hline
    Method & Accuracy & Macro-F1\\
    \hline
    SSL-ECG~\cite{sarkar2020self} & $93.7$ & $89.2$ \\
    SimCLR~\cite{chen2020improved} & $96.0$ & $93.5$\\
    TS-TCC~\cite{eldele2021time} & $97.2$ & $95.5$ \\
    AdaCT-T (124M) & $97.4$ & $96.5$ \\
    AdaCT-T (355M) & $\mathbf{98.7}$ & $\mathbf{97.9}$\\
    \hline
\end{tabular}
\caption{Comparisons of our AdaCT-T with those of the previous state-of-the-art self-supervised learning methods on Epilepsy Seizure Prediction.}
\label{tab1}
\end{table}

\begin{table}[b]
\centering
\begin{tabular}{|l|l|l|}
    \hline
    Method & Accuracy & Macro-F1\\
    \hline
    SSL-ECG~\cite{sarkar2020self} & $74.6$ & $65.4$ \\
    SimCLR~\cite{chen2020improved} & $78.9$ & $68.6$\\
    TS-TCC~\cite{eldele2021time} & $83.0$ & $73.6$ \\
    AdaCT-I (28M) & $83.4$ & $74.5$  \\
    AdaCT-I (87M) & $\mathbf{86.3}$ & $76.4$ \\
    \hline
\end{tabular}
\caption{Comparisons of our AdaCT-I with those of the previous state-of-the-art methods on SLEEP-EDF.}
\label{tab2}
\end{table}

\begin{table}
\centering
\begin{tabular}{|l|l|l|}
    \hline
    Method & Accuracy & Macro-F1 \\
    \hline
    SSL-ECG~\cite{sarkar2020self} & $65.3$ & $63.8$ \\
    SimCLR~\cite{chen2020improved} & $81.0$ & $80.2$ \\
    TS-TCC~\cite{eldele2021time} & $90.4$ & $90.4$  \\
    AdaCT-I (28M) & $98.2$ & $98.1$  \\
    AdaCT-I (87M) & $\mathbf{9 9 . 6 }$ & $\mathbf{9 9 . 5 }$  \\
    \hline
\end{tabular}
\caption{Comparisons of our AdaCT-I with those of the previous state-of-the-art methods on UCI HAR.}
\label{tab3}
\end{table}

\subsection{Baseline Comparison}
We select SSL-ECG~\cite{sarkar2020self}, SimCLR~\cite{chen2020improved}, and TS-TCC~\cite{eldele2021time} as baseline methods, previously evaluated on the same datasets by ~\cite{eldele2021time}. Our experiments follow identical conditions in terms of dataset partitioning and evaluation metrics, ensuring a fair comparison.

\subsubsection{AdaCT-T on Epileptic Seizure Prediction}

Given that the Epileptic Seizure Dataset consists of single-channel EEG data, with each chunk containing only 178 data points, we select AdaCT-T to convert input EEG data into string text. AdaCT-T enables us to leverage language transformer models as feature extractors for the converted text. We utilize GPT-2~\cite{radford2019language} models of 124M and 355M parameters, pre-trained on large-scale corpora using self-supervised learning. Table \ref{tab1} presents a comparison of our method with baseline approaches on the Epileptic Seizure Prediction dataset.

The experimental results indicate that the proposed method surpasses other state-of-the-art approaches. Specifically, in leveraging self-supervised learning models, the advantage of GPT-2 lies in its ability to effectively capture complex semantic relationships and patterns within the textual representation of EEG data. This capability stems from GPT-2's architecture, which employs a transformer-based model trained on extensive text corpora using self-supervised learning techniques. By fine-tuning GPT-2 on the converted textual representation of EEG data, our method effectively harnesses the pre-learned linguistic knowledge encoded within the model, facilitating superior performance in EEG-related tasks such as prediction and classification. 

Regarding the conversion of raw EEG time series data into a text-based format suitable for input into GPT-2, AdaCT-T plays a pivotal role. AdaCT-T enables the seamless transformation of EEG data into coherent text strings, preserving the temporal dynamics and inherent structure of the original data. This transformation process involves scaling each temporal data point into a three-digit integer representation and applying non-overlapping sliding window downsampling to accommodate the maximum input length of the GPT-2 model. These preprocessing steps optimize the input format for subsequent model utilization, ensuring efficient extraction of meaningful features from the EEG data.

We also show that applying AdaCT to a larger pre-trained model achieves a better performance: +1.3\% for GPT-2 355M (98.7\%) over GPT-2 124M (97.4\%) by prediction accuracy. Inspired by this, we conduct comparative experiments for fine-tuning pre-trained models of different sizes and include them in the ablation study section.

\subsubsection{AdaCT-I on Sleep-EDF}

For the Sleep-EDF dataset, which comprises EEG data with each piece containing 3000 data points, we utilize the AdaCT-I method. AdaCT-I focuses on converting multi-channel or lengthy EEG data into spatio-temporal 2D pseudo-images, making it suitable for integration with pre-trained visual transformers.

We use Swin Transformer models as feature extractors for the converted pseudo-images. Specifically, we utilize the Swin-Transformer-v2 tiny model~\cite{DBLP:journals/corr/abs-2111-09883}, which consists of 28 million parameters and is pre-trained on ImageNet-1k~\cite{deng2009imagenet}, as well as the Swin-Transformer-v2 base model, which comprises 110 million parameters and is pre-trained on ImageNet-21k.
Table \ref{tab2} presents the comparison results between the baseline methods and ours on the Sleep-EDF dataset.

The experimental results demonstrate the efficacy of our proposed method. By utilizing pre-trained Swin Transformer models as the core feature extractors for the converted pseudo-images, we achieve superior performance compared to other state-of-the-art approaches. The Swin Transformer's ability to capture both local and global features from the input data enables effective feature extraction from the spatio-temporal 2D pseudo-images generated by AdaCT-I. Additionally, in the process of converting raw EEG time series data into RGB pseudo-images suitable for input into Swin Transformer, AdaCT-I plays a pivotal role. AdaCT-I seamlessly transforms multi-channel or lengthy EEG data into spatio-temporal 2D pseudo-images, ensuring compatibility with Swin Transformer models and facilitating effective feature extraction. This integrated approach leverages the strengths of both AdaCT-I and Swin Transformer models, therefore leading to superior performance compared to other state-of-the-art approaches.


It is worth noting that there is a limitation when applying the AdaCT-T method on the Sleep-EDF Dataset. The generated text exceeds three times the acceptable input length of the largest pre-trained model we use. Even using non-overlapping sliding windows for downsampling may result in significant information loss. For further details, we illustrate the superiority of choosing AdaCT-I on this dataset through comparative experimental results in the ablation study section.

\subsubsection{AdaCT-I on UCI HAR}

Although the UCI HAR dataset is not an EEG dataset, it serves as a time series prediction benchmark based on human body monitoring data for evaluating the effectiveness of our method in a broader range of tasks. Given the multi-channel nature of this dataset, we employ the AdaCT-I method, using the Swin Transformer-v2 tiny and base models, which are identical to the pre-trained visual transformers utilized for the Sleep-EDF dataset. 
Swin Transformer's superiority stems from its unique architecture, particularly its incorporation of shifted windows. This design allows for efficient capture of both local and global features, addressing potential information loss in EEG data processing. 

Table \ref{tab3} presents the comparative results between our method and baseline approaches on the UCI HAR dataset. Our method noticeably surpasses the previous state-of-the-art methods, achieving an improvement of +9.2\% in classification accuracy for AdaCT 87M (99.6\%) over TS-TCC (90.4\%), and an improvement of +7.8\% in classification accuracy for AdaCT 28M (98.2\%) over TS-TCC (90.4\%). Table \ref{tab3} also shows that AdaCT performs much better than the earlier models ~\cite{chen2020improved} and ~\cite{sarkar2020self}. 
The analysis of the epoch-wise training process, as illustrated in Fig. \ref{fig5}, reveals that AdaCT-I, when applied to various pre-trained transformers, surpasses the previous state-of-the-art method within five fine-tuning epochs.

\begin{figure*}
\centering
\includegraphics[width=0.8\columnwidth]{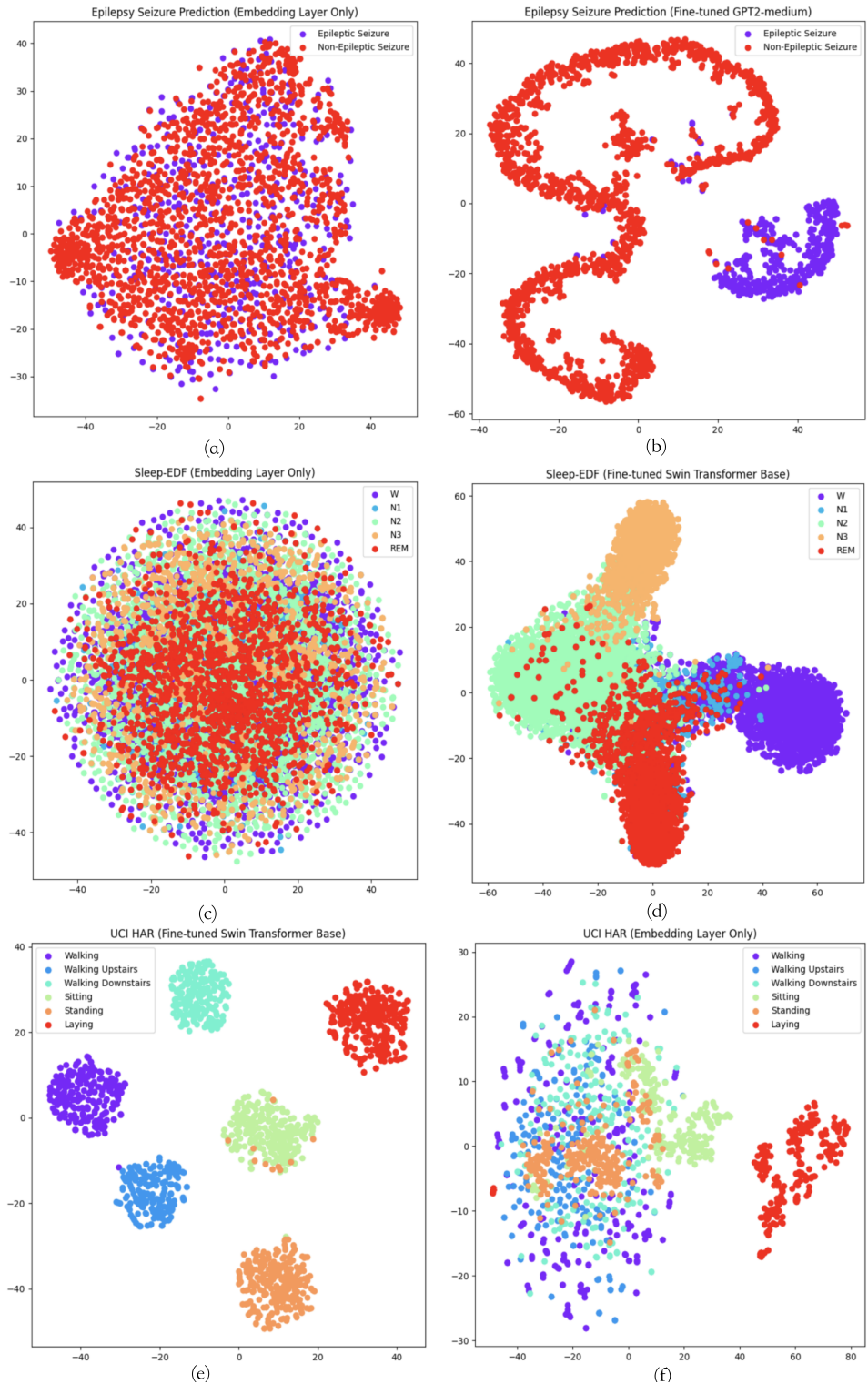} 
\caption{Visualization of Feature Embeddings from Fine-tuned Swin Transformer Models Compared to Embeddings from Embedding Layers. Each row represents a dataset, with the left side showing the embeddings from the embedding layer only, and the right side showing the embeddings from the last layer of our fine-tuned models. (a)-(b) correspond to Epileptic Seizure Prediction, (c)-(d) correspond to Sleep-EDF, and (e)-(f) correspond to UCI HAR.}
\label{fig6}
\end{figure*}

\subsection{Visualization}
The t-Distributed Stochastic Neighbor Embedding (t-SNE)~\cite{van2008visualizing} method is a powerful technique commonly used for visualizing high-dimensional data in lower dimensions while preserving local structure. In our study, we utilize t-SNE to compare the visualizations obtained from the embeddings of the first and last layers of our fine-tuned models on the test sets of the datasets mentioned above. The t-SNE plots reveal noteworthy differences between the two sets of embeddings, as shown in Fig. \ref{fig6}. While the embeddings from the first layer display scattered and overlapping clusters, indicating little  separation between classes, those from the last layer showcase clear and distinct clusters, with data points from different labels forming separate and well-defined groups. This disparity underscores the effectiveness of the feature extraction process in the deeper layers of our fine-tuned model, particularly in capturing label-related features essential for classification performance. 

\begin{table*}[t]
\centering
\resizebox{0.8\textwidth}{!}{
\begin{tabular}{|l|cc|cc|}
    \hline Approach (Pre-trained Model) & \multicolumn{2}{|c|}{ HAR } & \multicolumn{2}{c|}{ Sleep-EDF }  \\
    \cline{2-5}  & Accuracy & Macro-F1 & Accuracy & Macro-F1  \\
    \hline
    AdaCT-I-DeiT (86M)& $99.5 $ & $99.3$  & $80.7 $ & $70.3$ \\
    AdaCT-I-Swin-Transformer (87M) & $\mathbf{99.6}$ & $\mathbf{99.5}$  & $\mathbf{84.7}$ & $\mathbf{76.2}$ \\
    AdaCT-I-ViT (87M) & $99.5$ & $99.2$  & $79.9$ & $68.7$ \\
    \hline
    AdaCT-T-BERT (110M) & $ $ & $ $  & $77.0$ & $65.9$ \\
    AdaCT-T-GPT-2 (124M) & $ $ & $ $  & $76.6$ & $67.6$ \\
    \hline
    TS-to-Text-BERT (110M) & $83.3$ & $84.6$  & $71.1$ & $60.8$ \\
    TS-to-Text-GPT-2 (124M) & $85.5$ & $86.6$  & $72.0$ & $60.3$\\
    \hline
\end{tabular}
}
\caption{Comparisons of the proposed AdaCT-I, AdaCT-T and TS-to-Text on UCI HAR and SLEEP-EDF. Note: The blank cells in the table for HAR indicate that we do not design AdaCT-T method for multi-channel EEG data.
}
\label{tab4}
\end{table*}

\begin{figure*}
\centering
\includegraphics[width=1\columnwidth]{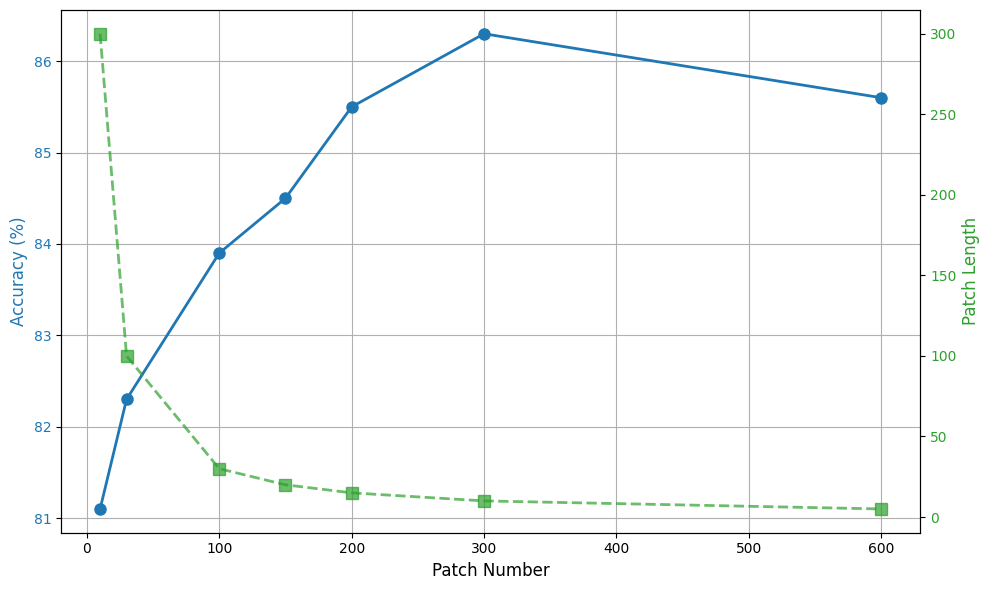} 
\caption{Parameter sensitivity experiments on the patch number of AdaCT-I's spatio-temporal reshaping step.}
\label{fig7}
\end{figure*}

\subsection{Ablation Study and Sensitivity Analysis}

We conduct ablation studies and sensitivity analysis on the UCI HAR Dataset and the Sleep-EDF Dataset.

To showcase the efficacy of AdaCT-T, we first design a control group named TS-to-Text. In this group, we use the conventional approach of directly inputting single-channel or multi-channel EEG data into language transformers and fine-tune them to evaluate their performance. Then, while keeping all parameters consistent in both pre-training and fine-tuning phases, we apply AdaCT-T to the same dataset and models. This allows for a direct comparison of the performance improvements achieved by AdaCT-T over the previous TS-to-Text approach. Table \ref{tab4} shows the performance improvements achieved by AdaCT-T over the conventional TS-to-Text approach on the Sleep-EDF Dataset.

Under the premise of maintaining consistent data set divisions and evaluation criteria, we further examine the effectiveness of utilizing AdaCT-I. Table \ref{tab4} illustrates that AdaCT-I achieves superior performance despite employing models with even fewer parameters. The performance gap is particularly evident on the UCI HAR Dataset, which further highlights the effectiveness of the adapter for extracting features from multi-channel time series signals.

To assess the influence of AdaCT-I's temporal segmentation ratios on prediction performance, we conduct sensitivity analysis experiments on the Sleep-EDF Dataset. These experiments aim to explore the sensitivity of the hyperparameter 'patch number', representing the number of patches obtained after temporal segmentation.  
From the results shown in Fig. \ref{fig7}, we can observe that our method achieves optimal performance on this dataset when the number of segments reaches 300. This underscores the effectiveness of the spatial-temporal segmentation and reshaping step. However, further increasing the segmentation beyond this point leads to overly dense partitioning, which disrupts the local temporal texture and consequently reduces accuracy.

We proceed to compare the performance of pre-trained vision transformers of varying sizes on the UCI HAR Dataset, as detailed in Table \ref{tab5}.
From the results we can find that fine-tuning a larger pre-trained model leads to a higher accuracy, indicating that AdaCT-I has the potential to be applied to larger models to tackle more complex time series decoding tasks. 

In fact, beyond the three datasets highlighted in the paper, we also evaluate the effectiveness of AdaCT-I on more public time series decoding datasets. Our proposed method demonstrates strong performance, for example, achieving classification accuracy exceeding 99\% on both BCI Competition IV dataset 1 ~\cite{van2008visualizing} and the Multimodal-Parkinson dataset ~\cite{zhang2022multimodal}. Due to the partial absence of baseline implementation details for these datasets, we do not present the results in tabular form.

\begin{table}
\centering
\resizebox{0.6\columnwidth}{!}{
\begin{tabular}{|l|l|l|}
    \hline
    Pre-trained Model & Accuracy & Macro-F1\\
    \hline
    AdacT-I-DeiT (22M) & $98.7$ & $97.6$  \\
    AdacT-I-Swin-Transformer (28M) & $98.2$ & $98.1$ \\
    AdacT-I-ViT (22M) & $98.8$ & $98.0$  \\
    \hline
    AdacT-I-DeiT (86M) & $99.5 $ & $99.3$  \\
    AdacT-I-Swin-Transformer (87M) & $\mathbf{9 9 . 6 }$ & $\mathbf{9 9 . 5 }$ \\   
    AdacT-I-ViT (87M) & $99.5$ & $99.2$  \\
    \hline
\end{tabular}
}
\caption{Comparisons of fine-tuning pre-trained vision transformers in different sizes on UCI HAR with our AdaCT-I.}
\label{tab5}
\end{table}

\section{Conclusion}\label{sec:con}
In conclusion, we introduce plug-and-play adapters designed to convert time series data into spatio-temporal 2D pseudo-images or text formats, further establishing a transfer learning framework for fine-tuning pre-trained vision or language transformers. The experimental results not only demonstrate the effectiveness of our proposed method but also illustrate its superior performance on larger-scale transformer models. Overall, our research provides novel insights into exploring the generalization potential of pre-trained large models in EEG and other time-series decoding tasks.

\section*{Acknowledgments}
This work is supported by the Project of Hetao Shenzhen-HKUST Innovation Cooperation Zone HZQBKCZYB-2020083 and Shenzhen Science and Technology Program (No. KQTD20180411143338837).

\bibliographystyle{unsrt}  
\bibliography{main}

\end{document}